# Adsorption, desorption, and interdiffusion in atomic layer epitaxy of CdTe and CdZnTe

E. M. Larramendi and O. de Melo[a]
*Physics Faculty, University of Havana10400 La Habana, Cuba*

M. Hernández Vélez
*Applied Physics Department, C-XII—Universidad Autónoma de Madrid and Porous Materials Group, Instituto de Ciencia de Materiales de Madrid—Consejo Superior de Investigaciones Científicas, Spain*

M. C. Tamargo
*Department of Chemistry, The City College of New York, New York, New York*

The mechanisms controlling the growth rate and composition of epitaxial CdTe and $Cd_{1-x}Zn_xTe$ films were studied. The films were grown by isothermal closed space configuration technique. A GaAs(100) substrate was exposed sequentially to the elemental sources (Zn, Te, and Cd) in isothermal conditions. While growth of ZnTe followed an atomic layer epitaxy (ALE) regime [self-regulated at one monolayer (ML) per cycle]; the CdTe films revealed different growth rates in dependence of the growth parameters (exposure and purge times). Combination of short purge times and larger Cd exposure times led to not self-regulated growth regime for CdTe. This is ascribed to large Cd coverages that were dependent on Cd exposure times (following a Brunauer-Emmett and Teller-type adsorption). However, for longer purge times and/or short Cd exposure times, an ALE self-regulated regime was achieved with 2 ML/cycle. In this sense, the self-regulation of the growth is limited by desorption, instead of absorption, as in the traditional growth technique. Cd atoms substitution by Zn atoms and subsequent evaporation of surface Cd atoms during Zn exposure has been proved. The influence of these facts on the growth and composition of the alloy is discussed. © *2004 American Institute of Physics*. [DOI: 10.1063/1.1813624]

## I. INTRODUCTION

Recent developments in optoelectronics have driven an extensive effort to grow II-VI compounds on various substrates.[1] CdTe and $Cd_{1-x}Zn_xTe$ (CZT) bulk crystals and epilayers have received considerable attention due to their advantages as substrates for the growth of HgCdTe and for the applications in optoelectronic devices. For example, CZT with band-gap energy ranging 1.5–2.4 eV is a promising candidate for tandem solar cell[2] and low-dimensional systems.[3] High-quality structures based on CdTe and CZT films have been grown by molecular beam epitaxy (MBE),[4] metal-organic chemical-vapor deposition,[3,5,6] and ultrahigh vacuum atomic layer epitaxy (UHV-ALE).[7,8] In ALE regime, the surface is alternately saturated with anion and cation fluxes with an adjustable dead time between successive exposures. At adequate temperatures, it is a self-regulated growth process with a growth rate (ML/cycle) governed by the surface coverage.[9] Recently, epitaxial films of ZnTe (Ref. 10), CdTe (Ref. 11), and CZT (Ref. 12) have been obtained by a technique (ICSSE, isothermal closed-space sublimation and epitaxy) based on ALE procedure using an isothermal closed space configuration. While for the growth of ZnTe films by ICSSE a 1 ML/cycle ALE regime was obtained, in the case of CdTe, growth rates larger than 1 ML/cycle and dependent of the exposure times were achieved at different growth conditions. The aim of this work was to investigate the influence of the surface stability, associated to different growth kinetics processes, on the growth rate regulation of CdTe and on the composition and thickness of CZT films.

## II. EXPERIMENTAL DETAILS

The samples were grown using a method similar to the traditional closed space sublimation method; its characteristics have been reported elsewhere.[10–12] A GaAs substrate is exposed alternately to the elemental sources. The whole system is at the same temperature. For that reason, one would expect that the only driving force for the film growth is the difference in chemical potential (vapor pressure) between the pure elements at the sources and the growing surface. The GaAs(100)-oriented substrates were previously chemically etched in a $H_2SO_4:H_2O_2:H_2O$ (5:1:1) solution for 60 s, in HF for 30 s, and thoroughly rinsed in bidistilled water. The growth temperature was 385 °C and the process occurred in a palladium-purified $H_2$ flux at atmospheric pressure. Between consecutive exposures to the elemental sources, the substrate is exposed for a certain period to the purified hydrogen environment in order to remove the residual vapors from the substrate cavity to avoid intermixing of vapors of different elements. This time is called purge time. Different sets of purge (3, 10, and 20 s) and exposure (3, 6, and 10 s) times were used to evaluate the influence of the growth parameters in the ICSSE growth of CdTe. This would bring out samples with different thickness and growth rates. The CZT films were grown using repetitions of a cycle basic sequence (period) composed by simple cycles of the binary compounds. Every basic sequence was then characterized by the number of simple cycles of CdTe ($n$) and ZnTe ($m$). For example, the exposure sequence Cd–Te–Cd–Te–Zn–Te–Cd–Te–Cd–Te–Zn–Te… has $n=2$ cycles of CdTe and $m=1$ cycle of ZnTe in the basic sequence. This basic sequence was repeated up to the completion of a total amount of 120 simple cycles (of binary compound) in all cases. The achievement of a superlattice is not expected because of the high growth temperature; instead, a homogeneous alloy was grown. Several combinations of $n$ and $m$ were used in order to obtain films with different cation molar fractions (see Table I).

---
[a] Electronic mail: omelo@fisica.uh.cu

TABLE I. Cycle sequences, thickness, and molar fraction for CZT samples. All the samples have 120 simple cycles and were left 10 s over the sources and 3 s in the purge hole. The samples with * were left 10 s in the purge hole.

| Sample | $m$ (ZnTe) | $n$ (CdTe) | Measured thickness (nm) | Predicted thickness (nm) | Measured Zn molar fraction | Predicted Zn molar fraction |
| --- | --- | --- | --- | --- | --- | --- |
| C75 | 0 | 1 | 180 | 214 | 0 | 0 |
| C32 | 1 | 2 | 159 | 155 | 0.11 | 0.08 |
| C41 | 2 | 2 | 100 | 125 | 0.14 | 0.15 |
| C38 | 1 | 3 | 129 | 170 | 0.18 | 0.05 |
| C58 | 3 | 3 | 72 | 125 | 0.24 | 0.15 |
| C39 | 2 | 3 | 124 | 143 | 0.62 | 0.11 |
| C20 | 1 | 1 | 120 | 125 | 0.64 | 0.15 |
| C33 | 2 | 1 | 109 | 96 | 0.84 | 0.27 |
| C56 | 3 | 2 | 66 | 107 | 0.85 | 0.21 |
| C46 | 3 | 1 | 89 | 81 | 0.87 | 0.35 |
| C9 | 1 | 0 | 36 | 36 | 1 | 1 |
| C69* | 1 | 2 | 122 | 103 | 0.45 | 0.13 |
| C70* | 1 | 3 | 129 | 111 | 0.46 | 0.08 |

The GaAs(100) substrates were previously chemically etched in a $H_2SO_4:H_2O_2:H_2O$ (5:1:1) solution for 60 s; in HF for 30 s, and thoroughly rinsed in bidistilled water. X-ray diffraction patterns were obtained in a Siemens D500 diffractometer using Cu-$K_\alpha$ radiation. Reflectivity measurements were performed using an UV-VIS Unicam photospectrometer. The thickness of the films was calculated from the reflectivity data on the basis of a simplified model of the interband transitions (model dielectric function) including interference. The details of this model have been published previously.[13] Growth rates (ML/cycle) were calculated from the slope of the plots of thickness as a function of cycles.

## III. RESULTS AND DISCUSSION

The good crystalline quality of epitaxial ZnTe, CdTe, and $Cd_{1-x}Zn_xTe$ thin films grown in our isothermal closed space configuration system was confirmed by x-ray diffraction and transmission electron microscopy. The results of these studies were presented in previous papers.[10–12] In the case of ZnTe, both exposition and purge times were 3 s and the growth temperature was 385 °C. These times were enough to deposit 1 ML/cycle [1 ML of ZnTe(100) =0.305 nm]. Large source exposure times were also used but no modification of the growth rate was detected[10] as expected for a 1 ML/cycle self-regulated ALE regime. On the other hand, different growth rates (ML/cycle) were calculated for CdTe samples grown with different sets of exposure and purge times. In Fig. 1, three different growth rates are observed, 5.5, 3.5, and 2 ML/cycle [1 ML of CdTe(100) =0.324 nm]. It can be noted that the combination of the largest Cd exposure time and the shortest purge time in the set 10-3-10 (in seconds, Cd exposure time—purge time—Te exposure time) led to the higher growth rate ($\sim$5.5 ML/cycle). In this set of samples it is also observed that the best lineal fitting of the experimental data does (dotted line in Fig. 1) not intercept the time axis at zero. This fact will be commented after since it is not relevant for the present discussion. The following sets to be analyzed are sets 6-3-6 and 10-10-10. They represent an intermediate condition in which the source exposure time (purge time) was decreased (increased) with respect to the set 10-3-10. It is observed that both effects tend to decrease the growth rate to an intermediate value of 3.5 ML/cycle. Finally, the sets with a larger relation between purge and exposure times (sets 3-10-3, 3-10-10, and 10-20-10) display the lower growth rate of 2 ML/cycle, and the variation of the exposure time to Te does not affect this result.

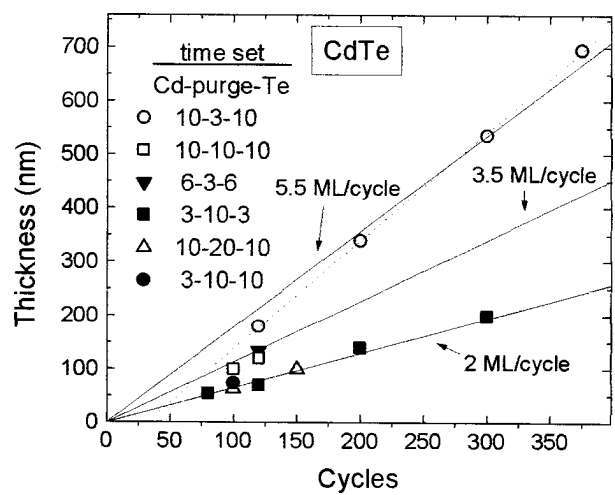

FIG. 1. Thickness dependence on the number of cycles for CdTe samples in different growth conditions. The used time sets are shown with the following order: Cd exposure time—purge time—Te exposure time.

Growth rates higher than 2 ML/cycle can be ascribed to the combined effect of (i) not self-regulated Cd-adsorbed surface coverage larger than 2 ML during the Cd exposure step and (ii) no completion of Cd desorption during purge time for short purge times. If for larger purge times desorption saturates at a stable Cd surface, this will be a self-regulation mechanism for the growth of CdTe. This condition seems to be attained for the time combinations when the relationship between the purge time and the Cd exposure time is large (sets 3-10-3, 3-10-10, and 10-20-10). All these sets presented the same growth rate of 2 ML/cycle (growth rates lower than this value were never obtained for our CdTe samples). This seems to indicate that a surface coverage of 2 ML of Cd is the stable one under a flow of hydrogen at 385 °C. Then, large purge times and reasonably short exposure times allowed the self-regulation of the growth of CdTe leading to an ALE regime with 2 ML/cycle. ALE growth rates larger than 1 ML/cycle for CdTe, but due to large Te coverage, have been detected in MBE experiments.[9] Contrary to the traditional nonequilibrium growth techniques like MBE and chemical vapor deposition (CVD), in ICSSE, the self-regulated condition is obtained by desorption saturation instead of adsorption saturation. This can be due to the variation of the thermodynamic configuration in ICSSE: semi-close configuration during exposure steps and open configuration during purging steps.

The not regulated Cd adsorption during Cd exposure (in spite of the isothermal regime in which source and growing surface are at the same temperature) can be explained in the following way. As the growth temperature (385 °C) is higher than the Cd melting point, evaporation instead of sublimation takes place in the Cd source. Then, the equilibrium over the Cd source occurs among the liquid Cd, the vapor Cd, and the Cd at the solid surface. The equilibrium will be attained when the vapor pressure of Cd (in this case equal to the equilibrium vapor pressure of liquid Cd) will be equal to the equilibrium vapor pressure of the Cd-covered surface. This can occur only with liquid Cd in the surface; however, we did not observe any evidence of liquid in any step of the layer formation. Then, the equilibrium vapor pressure of the solid Cd-covered surface is lower than that of the liquid Cd source and the coverage of the surface after a single exposure to Cd will be of several monolayers and will depend on the exposure time. This effect is similar to that called bulk condensation according to Brunauer-Emmett and Teller-(BET) type adsorption.[14] When placing the surface on the purge hole after the Cd exposure, part of the adsorbed Cd is desorbed until the surface stability is attained or the purge time finishes. Over the Te source, most of the Cd incorporates to the layer. Te source sublimates and it is expected that the adsorption of this element is regulated or controlled by the equilibrium between the vapor pressure of the growth surface and that of the solid source. Therefore, when exposing to Te, the surface Cd coverage will determine the growth rate. The main parameter for determining the growth regime is the relationship between the purge and the exposure time to Cd.

In the ZnTe case,[10] the partial vapor pressure difference between the film surface and the solid source could be neglected because there is not any liquid source. Then, a self-regulated adsorption mechanism takes place over the solid sources and the stability of the surfaces is reached immediately over the purge hole. The nonregulated regime would be used to grow CdTe films when thicker films are desired, and a precise control of the thickness is not so necessary. In nonequilibrium growth techniques like MBE and CVD, the bulk condensation of the species on the growth surface never occurs even if a source is in the liquid phase. This is because in such techniques, the beam equivalent pressure in the chamber never attains the equilibrium vapor pressure of the source.

The non-self-regulated growth of CdTe is expected to be sensitive to changes in the kinetic parameters during the growth. It is known that desorption and adsorption rates depend on the configuration and morphology of the surface. For example, the lattice strain reduces the cation incorporation rate by lowering the energy barrier for thermal desorption.[15] Therefore, it is expected that the growth rate will not be constant during the heterostructure growth, using a non-self-regulated regimen. The growth rate rather varies in dependence on the characteristics of the surface. From Fig. 1, it can be noted that the samples, grown with the time combination 10-3-10, agree with a lineal dependence (~6.2 ML/cycle, dot line), which does not intercept the time axis in zero. This can be due to a lower growth rate due to the layer strain at the initial stage of the heterostructure growth. For that reason, the growth in not self-regulated regime is not only determined by the time combination; it is also determined by the surface condition. It could be noticed that for the samples with a growth rate of 2 ML/cycle, the lineal fitting to the experimental data goes by zero, in correspondence with a constant growth rate due to a self-regulated regime. In fact, this also supports the assumption of such a regime at these conditions.

Table I shows the results of the Zn molar fraction of CZT layers determined by x-ray diffraction. The expected composition of the alloys can be estimated taking into account the growth rates of the binary compounds 1 ML/cycle for the ZnTe and 5.5 or 3.5 ML/cycle for the CdTe (according to the time combination used). From Table I, it is observed that the measured Zn molar fraction exceeds the estimated one. This could suggest substitution of Cd atoms by Zn atoms during the Zn exposure. In order to confirm this hypothesis, it was carried out in the following experiment: Two samples were grown according to the sequences: (A) Te–Zn–Cd–Te–Zn–Cd… and (B) Te–Cd–Zn–Te–Cd–Zn… For the sample A (B) it would be expected that the Cd (Zn) will not be deposited on a Zn (Cd) covered surface, since both elements share the same place in the crystalline alloy lattice. For this reason, the sample A would be pure ZnTe and the sample B pure CdTe. Figure 2 shows the results of the energy dispersive spectroscopy (EDS) for both samples. In the sample B, the presence of Zn was detected, while in A any evidence of Cd was not detected. This means that the atoms of Zn incorporate into a surface of the CZT alloy terminated in Cd atoms, while the atoms of Cd do not incor-porate into a surface of the CZT alloy terminated in Zn atoms. Also, from both results it can be deduced that the Zn incorporation is necessarily for the substitution of the Cd atoms. This conclusion could be related with the fact that the Zn–Te chemical bond is more stable than the Cd–Te one.[15] This substitution process among cations does not affect the amount of deposited material. However, it can produce great Zn incorporations into the alloys during the Zn exposure, since the Cd atoms of more inner layers can interdiffuse toward the surface and be substituted by Zn atoms as well. A detailed analysis of the cation substitution process in MBE CdZnTe will be presented in Ref. 16. Other effects could affect the alloy composition of the samples obtained by this procedure. In fact, it is interesting to note that the alloy layers seem to "prefer" some composition values (for instance, 0.85, 0.63, and 0.45). This can be associated with thermodynamic equilibrium at energy minima for these compositions of the alloy. However, we have not found evidences of superstructure related to the order in the x-ray or electron-diffraction patterns of these samples. Further investigations are needed in this topic.

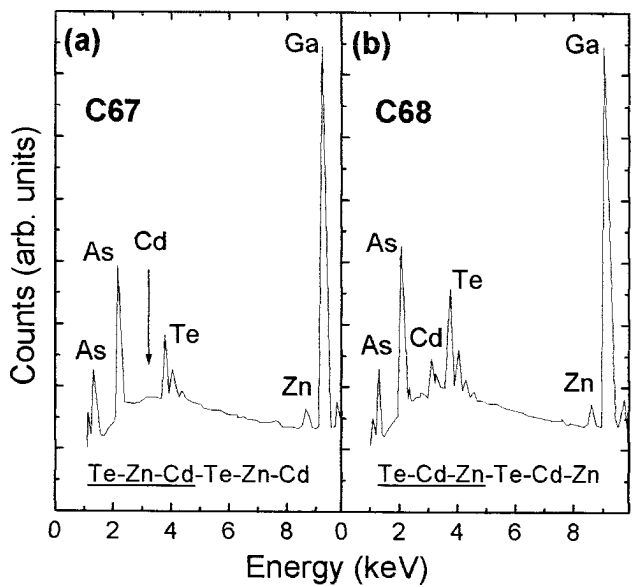

FIG. 2. Energy dispersive spectra for samples (a) sample A (Te–Zn–Cd… exposure sequence) and (b) sample B (Te–Cd–Zn… exposure sequence).

A similar analysis can be carried out in relation with the alloy thickness. In Table I the results of the measured and estimated thickness (according to the growth rates of the binary compounds) are shown. It can be noted that there is not always a good agreement. Instead, most of the values of expected thickness exceed the results of the measurements. This is because the growth of CdTe takes place in a non-self-regulated regime, where the growth rate (ML/cycle) depends mainly of the surface condition in each cycle of CdTe. As was evidenced above, the growth rate of CdTe in a non-self-regulated regime is smaller at the initial stages of the CdTe/GaAs growth. In the case of the alloy growth, the surface condition can be affected by the proximity of the interfaces and/or by the alternate deposition of different materials. Consequently, it is not appropriate to estimate the alloy thickness using the average growth rate obtained for thick layers of CdTe. The growth rate of the alloy in a non-self-regulated regime depends on many factors, among them are the layer strain, the cations substitution and interdiffusion processes, and the composition and morphology of the growing surface.

## IV. CONCLUSIONS

The main characteristics of the CdTe growth by using isothermal closed space sublimation and Epitaxy have been analyzed. The liquid Cd source leads to the non-self-regulated regime in CdTe growth. This is because a high Cd coverage is consistent with a BET-type adsorption. However, longer purge time favors the Cd desorption, leading to a stable Cd coverage and enabling an ALE self-regulated regime with 2 ML/cycle. The control of thickness and composition in the digital CdZnTe alloy is a very difficult task, since that it will depend on several factors, as the morphology, the interdiffusion of Zn and Cd, or the strain of the structures.

## ACKNOWLEDGMENTS

The authors acknowledge the support of Alma Mater grant 1002 from the University of Havana and the assistance of A. Garcia and S. Tobeñas. F. Fernámdez-Lima and E. Pedrero are acknowledged for the EDS measurements. One of the authors (M.C.T.) thanks the Visiting Consulting Professor Program of the International Center for Theoretical Physics.

[1] S. J. C. Irvine, A. Stafford, and M. U. Ahmed, J. Cryst. Growth **197**, 616 (1999).
[2] T. L. Chu, S. S. Chu, C. Ferekides, and J. Britt, J. Appl. Phys. **71**, 5635 (1992).
[3] E. J. Mayer, N. Pelekanos, J. Khul, N. Magnea, and H. Mariette, Phys. Rev. B **51**, 17263 (1995).
[4] R. Miles, G. Wu, M. Johnson, T. McGill, J. Faurie, and S. Sivananthan, Appl. Phys. Lett. **48**, 1383 (1986).
[5] M. Levy, N. Amir, E. Khanin, A. Muranevich, Y. Nemirovky, and R. Beserman, J. Cryst. Growth **187**, 367 (1998).
[6] K. Cohen, S. Stolyarova, N. Amir, A. Chack, R. Beserman, R. Weil, and Y. Nemirovky, J. Cryst. Growth **198**, 1174 (1999).
[7] J. Sadowski and M. Herman, Appl. Surf. Sci. **112**, 148 (1997).
[8] J. Sadowski and M. Herman, Thin Solid Films **306**, 266 (1997).
[9] B. Daudin, D. Brun-Le Cunff, and S. Tatarenko, Surf. Sci. **99**, 352 (1996).
[10] E. M. Larramendi et al., J. Cryst. Growth **223**, 447 (2001).
[11] E. M. Larramendi, E. Purón, and O. de Melo, Phys. Status Solidi B **230**, 339 (2002).
[12] S. Tobeñas, E. M. Larramendi, E. Purón, O. de Melo, F. Cruz-Gandarilla, M. Hesiquio-Garduño, and M. Tamura, J. Cryst. Growth **234**, 311 (2002).
[13] E. M. Larramendi, E. Purón, and O. de Melo, Semicond. Sci. Technol. **17**, 8 (2002).
[14] S. J. Gregg and K. S. W. Sing, *Adsorption, surface area, and porosity* (Academic, London, 1982), p. 42.
[15] M. A. Herman, A. V. Kozhukhov, and J. T. Sadowski, J. Cryst. Growth **174**, 768 (1997).
[16] E. M. Larramendi, O. de Melo, and I. Hernández-Calderón, J. Appl. Phys. (to be published).